\documentclass[preprint,showpacs,amsmath,amssymb,floatfix,fleqn,aps]{revtex4}
\usepackage{graphicx}
\usepackage{dcolumn}
\usepackage{bm}
\begin{document}

\title{Effective interactions in the colloidal suspensions from HNC theory}

\author{Daniel L\'eger}
\affiliation{Laboratoire de M\'ecanique et d'Energ\'etique
Universit\'{e} de Valenciennes \\ et du Hainaut-Cambr\'esis
Le Mont Houy, 59313 Valenciennes Cedex 9, France.}
 \email{daniel.leger@univ-valenciennes.fr}
\author{Dominique Levesque ${}^{1}$}
\affiliation{Laboratoire de Physique Th\'eorique  UMR CNRS 8627 
 Universit\'{e} Paris Sud B\^atiment 210,   
91405 Orsay, France.}
\email{dominique.levesque@th.u-psud.fr}

\begin{abstract}
 The HNC Ornstein-Zernike integral equations
 are used to determine the properties of simple models of colloidal solutions  
 where the colloids and ions are immersed in a solvent considered as a 
 dielectric continuum and have a size ratio equal to 80 and a charge ratio 
 varying between 1 and 4000. At an infinite dilution of colloids, 
 the effective interactions between colloids and ions
 are determined for ionic concentrations ranging from 0.001 to 0.1 mol/l 
 and compared to those derived from the Poisson-Boltzmann theory. 
 At finite concentrations, we discuss on the basis of the HNC 
 results the possibility of an unambiguous definition of the effective 
 interactions between the colloidal molecules.\par
\bigskip 
 {${}^1$To whom correspondence should be addressed}

\end{abstract}
\maketitle
\newpage 
                       
\section{Introduction}
The large difference of sizes and electric charges between the colloids
and  molecules or atoms present in the colloidal suspensions
is one of the main difficulties encountered in the theoretical study of these
multicomponent solutions. A customary simplification is to approximate the solvent contributions
to the solution properties  by those of a dielectric continuum 
where are immersed the colloids and ions. From this simplification, it results
that the colloidal suspensions are mixtures of colloids and ions 
where the coulombic interactions between charges are divided by the dielectric
constant of the continuum. All excluded volume and polarization effects
due to the solvent are neglected or assumed to increase notably the ion sizes by
considering that the ions are surrounded with a shell of bound molecules of solvent,
for instance a hydration shell. A new simplification is generally done since the colloids are 
considered as structureless, rigid, spherical and charged particles. With these approximations,
the estimate of properties of colloidal suspensions is reduced to that of mixtures 
of spherical ions and particles where the size and charge of ions and particles 
have highly different values \cite{Over,Hans,Lowen}.\par

 The screening effects which are present in the coulombic systems, 
lead to further simplify the description
of the colloidal suspension by supposing that the interaction between
colloidal particles due to their electric charges can be represented
by an effective potential resulting from the screening of their own charges 
by the ions \cite{Over,Derja,Louis,Louisa}. The standard methods of liquid theory allow to avoid this latter 
approximation and to calculate the structural order and thermodynamic properties of the
colloidal suspensions as those of mixtures of large charged particles and small ions. 
The main aim of this work is to demonstrate that the HNC Ornstein-Zernike (OZ) integral 
equations \cite{Macd} allow to perform such a calculation for a size ratio between spherical 
colloidal particles and spherical ions equal to 80 and charge ratios ranging from
1 to 4000. Although these charge ratios are currently found in real suspensions, 
the size ratio is small compared to that characteristic of real colloids.
However, a size ratio equal to 80 preserves the fact that the external surface
of colloids is sufficiently large for allowing that the screening of 
the colloid charge  by the ions is not hindered by excluded volume
effects, because the external colloid surface is able to accommodate about
25000 small ions. Such a size ratio seems sufficient to perform a valuable comparison 
with the Poisson-Boltzmann (PB) theoretical approaches of the estimate of the effective 
interaction between ions and colloids in the infinite dilution limit, in particular for 
values of the charge ratio larger than 1000.

In addition to the coulomb interactions, short range interactions of van der Waals
type exist between the ions and colloids. These interactions will be also discarded
in order to determine unambiguously the effective interactions 
induced solely by the interactions between the charges. The specific system studied
is a neutral mixture of large charged hard spheres, the colloids, soluted in a fluid
of one or two species of small charged hard spheres, the ions. The colloids are positively
charged. A link between this model system and the real suspensions is made
by attributing to the ions a hard core diameter of $\sigma_i = 5 $ \AA, a typical value  for
hydrated ions, giving for the colloid diameter a value of $\sigma_c = 400 $\ \AA  ; 
this equivalence allows to express the ionic densities of the model in mol/l (M).

The numerical procedure used to solve the HNC integral equations is described
in Section II, in the same section the results obtained at an infinite dilution
of colloids are given and compared to those of the PB theory. 
In Section III we present the results obtained
at finite concentration of the colloids. In the final Section, we summarize
the results and discuss possible extensions of this work.

\section{Infinite dilution}

To solve numerically the HNC OZ equations for symmetric and asymmetric 
mixtures of charged hard spheres, a sufficiently small integration step is needed 
to describe  accurately  the very steep peak existing 
at contact in the correlation functions between the negative ions and colloids
\cite{Bel,Khan,Vlac,Forc,Bell,Lai,Bello}. 
Furthermore, the upper bound of the numerical integrations must 
be equal to several colloid diameters in order to study
systems with a colloid finite concentration. Finally, at low ionic concentrations, 
the Debye length characterizing the screening effects can have a
value of the same order than the colloid diameter which requires also
the use of an upper bound of integration large enough to reach the
asymptotic behavior of the correlation functions.
With these constraints and the fact that the numerical solution
of the HNC equations is obtained using a Fast Fourier transform
algorithm, the numerical integrations are performed with $2^{18}=262144$
$\Delta r$ or $\Delta k$ steps in the $r$ and $k$ spaces. The colloid radius 
$R_c$, chosen as the unit of length, is equal to 16000 $\Delta r$ and
the ion radius to 200 $\Delta r$. In $r$  space, the integration upper bound
is equal to 16.38 $R_c$. 
\begin{figure}
\includegraphics{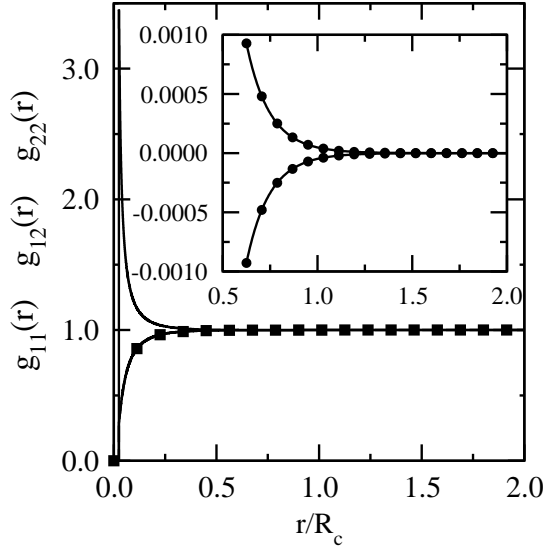}
\caption{\label{fig:epsart}  Ion-ion correlation functions $g_{11}(r)$, $g_{12}(r)$ 
(solid lines) and $g_{22}(r)$ (filled squares) with $z_1=-1$, $z_2=1$,
$\rho_v=0.01\, M$  and $\rho_0=0$. Insert : fit of $g_{11}(r)$ and $g_{12}(r)$
(solid lines) by $A_{\alpha\, \beta} \exp(-\kappa_p  r)/r$ 
(filled circles) ($\kappa_p R_c=6.62$).}
\end{figure}

 The coupling strength between charges is characterized by the Bjerrum length  
$l_B=e^2/\epsilon  k_B T$ where $\epsilon$ is the dielectric constant of the solvent,
$e$ the electron charge, $k_B$ the Boltzmann constant and $T$ the temperature.
For an aqueous suspension $l_B=7.198$ \AA \ at $T = 293$ K. The colloid and ion densities
$\rho_0$, $\rho_1$ and $\rho_2$ must be chosen in order to satisfy the electro-neutrality
relation $z_0\, \rho_0 + z_1\, \rho_1 + z_2\, \rho_2 \, = \, 0$ where $z_0$, $z_1$ and $z_2$
are the colloid and ion charges.  

\begin{figure}
\includegraphics{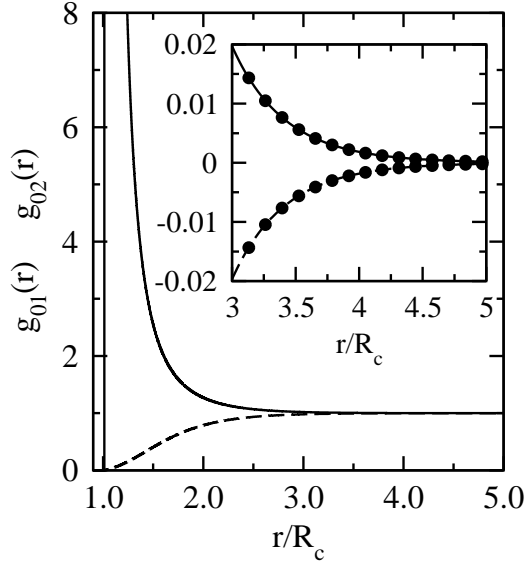}
\caption{\label{fig:epsarta} Ion-colloid correlation functions $g_{01}(r)$ 
(solid line) and $g_{02}(r)$ (dashed line) with $z_1=-1$, $z_2=1$,
$\rho_v=0.001\ M$  and $\rho_0=0$. Insert : fit of $g_{01}(r)$ (solid line)
and $g_{02}(r)$ (dashed line) by $A_{\alpha\, \beta} \exp(-\kappa_p  r)/r$ 
(filled circles) ($\kappa_p R_c=2.08$).}
\end{figure}

In the solution of the HNC equations, the long range of the coulomb potential is taken into account 
by writing the direct correlation functions $c_{\alpha \, \beta} (r)$  ($\alpha,\, \beta = 0,\, 1$ 
and $2$) :
\begin{eqnarray}
c_{\alpha \, \beta} (r) = a_{\alpha \, \beta} (a_l,r) - l_B \, z_\alpha \, z_\beta \,\frac{ u_l(a_l r)}
{r}
\end{eqnarray} 
where $u_l(a_l r)= {\rm erf}(a_l r)$ and $a_l$ is chosen equal to $0.9\, R_c$. The functions 
$a_{\alpha \, \beta} (a_l,r)$  have 
a short range, because the asymptotic behavior of $c_{\alpha \, \beta} (r)$ at large $r$ 
is equal to $-l_B \, z_\alpha \, z_\beta /r$. The Fourier transform ${\bar f}(k)$
of a function $f(r)$ being defined by
\begin{eqnarray}
{\bar f}(k) =  4 \pi \rho  \int_0^{\infty}  {\frac {\sin (kr)} {k r}} f(r) r^2 dr 
\end{eqnarray}
where $\rho = \rho_0 + \rho_1 +  \rho_2$ (for an infinite dilution of colloids $\rho_0 = 0$),
the Fourier transform of $u_l(a_l r)$ is
${\bar u}_{l}(k)=\exp (-k^2/(4 \, a^2_l))/k^2$ and that of $c_{\alpha \, \beta} (r)$
\begin{eqnarray}
{\bar c}_{\alpha \, \beta}(k) & = & {\bar a}_{\alpha \, \beta}(k)-c_l\, {\bar u}_{l}(k) \,
z_\alpha \, z_\beta \, .
 \label{dipola}
 \end{eqnarray} 
 where $c_l = 4 \pi \rho \, l_B$.\par
 In $k$-space, the OZ equations at a colloidal infinite dilution  write 
 for the ion correlation functions 
\begin{eqnarray}
{\bar h}_{11}(k) & = & \displaystyle{\frac{{\bar c}_{11}(k)+x_2({\bar c}_{12}^2(k)-{\bar c}_{11}(k)\,
{\bar c}_{22}(k))}{{\bar D}_0(k) }}\nonumber \\
{\bar h}_{22}(k) & = & \frac{{\bar c}_{22}(k)+x_1({\bar c}_{12}^2(k)-{\bar c}_{11}(k)\,
{\bar c}_{22}(k))}{{\bar D}_0(k) }\nonumber \\
{\bar h}_{12}(k) & = & \frac{{\bar c}_{12}(k)}{{\bar D}_0(k) }
 \label{dipolu}
\end{eqnarray}
and the ion-colloid correlation functions
\begin{eqnarray}
{\bar h}_{01}(k)  =  \frac{{\bar c}_{01}(k)\, (1-x_2\, {\bar c}_{22}(k))+x_2\, 
{\bar c}_{12}(k)\, {\bar c}_{02}(k)} {{\bar D}_0(k)} \nonumber \\
{\bar h}_{02}(k)  =  \frac{{\bar c}_{02}(k)\, (1-x_1 \, {\bar c}_{11}(k))+x_1 \,
{\bar c}_{12}(k)\, {\bar c}_{01}(k))} {{\bar D}_0(k)}
 \label{dipolb}
\end{eqnarray}
where $x_1=\rho_1/\rho$, $x_2=\rho_2/\rho$ and 
\begin{eqnarray}
{\bar D}_0(k)  =  (1-x_1\, {\bar c}_{11}(k))\, (1-x_2\, 
{\bar c}_{22}(k))-x_1\, x_2\, {\bar c}^2_{12}(k) 
 \label{dipolc}
\end{eqnarray}
supplemented by the HNC closure in $r$-space :
\begin{eqnarray}h_{\alpha \, \beta}(r)&&  =  g_{\alpha \, \beta} (r) -1\nonumber \\
&& =\exp (-l_B z_\alpha \, z_\beta /r
+h_{\alpha \, \beta} (r) -c_{\alpha \, \beta} (r)) -1 \nonumber \\ && = 
 \exp (-l_B z_\alpha \, z_\beta (1-{\rm erf}(a_l r))/r)\nonumber \\
&& \times \exp (h_{\alpha \, \beta} (r) -a_{\alpha \, \beta} (a_l,r)) -1\, .
 \label{dipole}
\end{eqnarray}
The convergence of the numerical solutions is obtained by using a Picard algorithm 
and considered to be achieved when
the two successive iterations $n$ and $n+1$ of the functions $\gamma_{\alpha \, \beta} (r)$ 
$=h_{\alpha \, \beta} (r) -a_{\alpha \, \beta} (a_l,r)$ satisfy the relations:
{\ \ \ \ \ \ \ \ }\begin{eqnarray}
{\ \ \ \ \ \ \ \ }\vert \gamma^{(n+1)}_{\alpha \, \beta} (r) \ - \ \gamma^{(n)}_{\alpha \, \beta} \, (r) \vert 
\ < \ 10^{-6} \, .
 \label{crit}
\end{eqnarray}

 \begin{figure}
\includegraphics{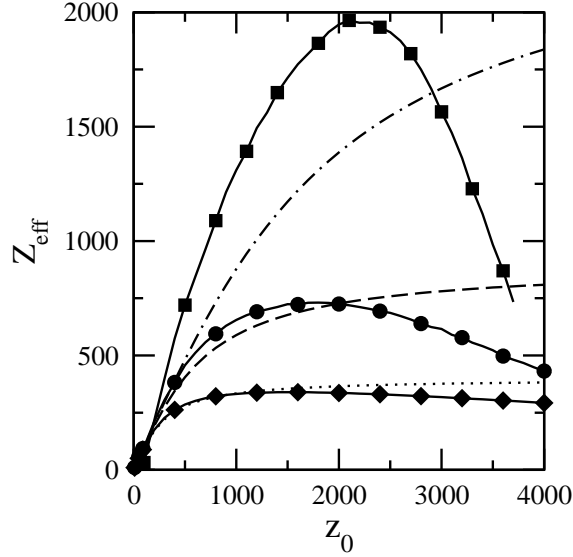}
\caption{\label{fig:epsartb} At $\rho_0=0$, PB and HNC estimates of $Z_{eff}$ versus $z_0$ 
for ionic charges $z_1=-1,\,z_2=1$ at $\rho_v=0.001$ M
 : PB dotted line, HNC solid line and filled diamonds, $\rho_v=0.01$ M
 : PB dashed line, HNC solid line and filled circles and $\rho_v=0.1$ M
 : PB dash-dotted line, HNC solid line and filled squares.}
\end{figure}

 \begin{figure}
\includegraphics{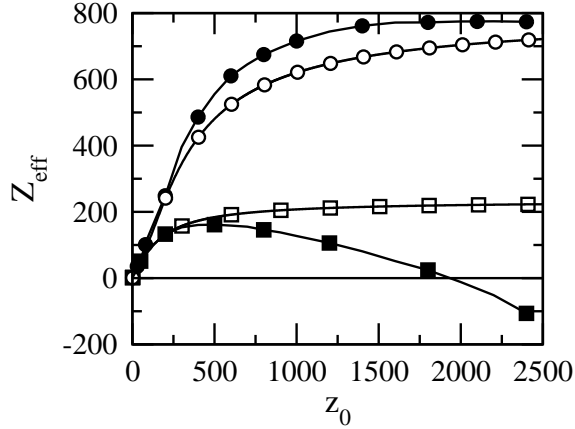}
\caption{\label{fig:epsartc}At $\rho_0=0$, PB and HNC estimates of $Z_{eff}$ versus $z_0$ 
at $\rho_v=0.001$ M, ionic charges $z_1=-1,\,z_2=2$,
 : PB solid line and white circles, HNC solid line and filled circles, 
and  $z_1=-2,\,z_2=1$  
 : PB solid line and white squares, HNC solid line and filled squares.}
\end{figure} 

 In the literature, detailed comparisons of the two-body correlation
 functions and thermodynamic properties
 computed from the HNC theory and Monte-Carlo simulations
 have been made for ionic solutions 
 \cite{Torr,Abra,Cacc,Kjell,Bes,Kalyuz,Uland,Lyu}. They have  defined the domain 
 of validity of the HNC approximation for ionic concentrations  
 $\simeq$ 1 M and charge and size ratios of the order of 1-20. These works have shown 
 that the HNC closure gives an accurate description of the correlations at large $r$ 
 in symmetric or asymmetric systems of charged hard spheres, 
 with equal or different diameters, immersed in a dielectric continuum. This satisfactory result 
 constitutes a favorable basis to the study of the infinite dilution of colloids in our model. 
 At $\rho_0 = 0$, the HNC equations (\ref{dipolu}) for the correlation functions
 of the small ions are independent of those giving the ion-colloid correlation functions 
 $g_{01} (r)$ and $g_{02} (r)$ (cf. Eqs. (\ref{dipolb})).  
 Hence, it is possible to determine $g_{01} (r)$ and $g_{02} (r)$ 
 at $\rho_0 =0$ for any charge $z_0$ when the equations (\ref{dipolu}) have been solved. 
 Such a computation has been realized for $z_0<4000$  and symmetric ($z_1=-1$, $z_2=1$)
 or asymmetric ($z_1=-2$, $z_2=1$) and ($z_1=-1$, $z_2=2$) ionic solutions
 where the densities $\rho_v$ of 
 the ionic species with the highest valence, are equal to $0.1$, $0.01$ and $0.001$ M
 (i.e. $\rho_v \sigma^3_i = 0.007528$,  $0.0007528$ and  $0.00007528$),
 respectively.\par 
  
\begin{figure}
\includegraphics{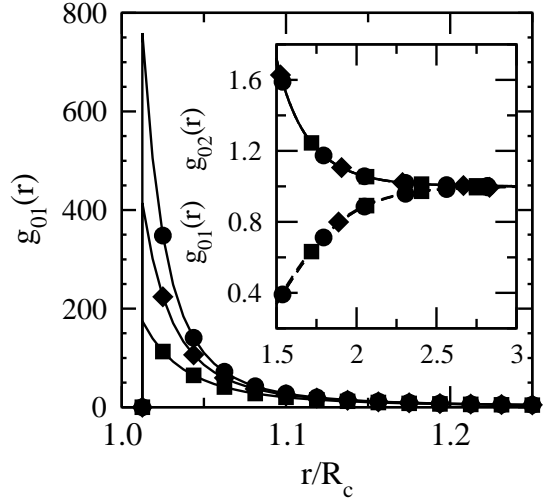}
\caption{\label{fig:epsartd} At $\rho_0=0$, $\rho_v=0.001$ M and ionic charges 
$z_1=-1,\,z_2=2$, variation of the functions $g_{01}(r)$ near
$r\simeq R_c$ : $z_0=1200$ solid line and filled squares,  
$z_0=1800$ solid line and filled diamonds,
$z_0=2400$ solid line and filled circles.
Insert : for $1.5R_c<r<3R_c$, $g_{01}(r)$ (solid line and
filled squares $z_0=1200$, diamonds $z_0=1800$ and circles $z_0=2400$) 
and $g_{02}(r)$ (dashed line and filled squares $z_0=1200$, diamonds 
$z_0=1800$ and circles $z_0=2400$).}
\end{figure}

  From the correlation functions, there are several possible definitions of the effective 
 interactions between colloids or more generally between specified components of a multi-component 
 mixture \cite{Hans,Louis}. One of those is the potential of mean force 
  $v^{m}_{\alpha \, \beta} (r)$ given by 
 $\log \{g_{\alpha \, \beta} (r)\}$ $=$ $ -\beta v^{m}_{\alpha \, \beta} (r)$. 
 Theoretical approaches, such as the DLVO theory \cite{Over,Derja}, provide an explicit form 
 for the effective interaction between colloids :
\begin{eqnarray}   
\beta  v^{eff}_{0\, 0} (r) = Z_{eff}^2 l_B \, 
\Bigg[{\frac  {\exp(\kappa_D R_c)} {  1+\kappa_D \, R_c} }\Bigg]^2 \,
{\frac {\exp(-\kappa_D \, r)}{r}}
\label{veff}
\end{eqnarray}
where $\kappa_D$ is the inverse Debye length defined 
by $\kappa^2_D=4 \pi l_B( z^2_1 \rho_1+ z^2_2 \rho_2)$,
and the parameter $Z_{eff}$ is conventionally called 
the effective charge of the colloids.\par

 In the  limit $\rho_0=0$, from the linearized PB theory,
a similar expression is obtained for the effective interaction between colloid and ions 
\begin{eqnarray}   
\beta  v^{eff}_{0\, \alpha} (r) = Z_{{eff}}  z_\alpha \, l_B \,
{\frac  {\exp(\kappa_D \,  R_c)} {(1+\kappa_D R_c)} }
{\frac  {\exp(-\kappa_D r)} { r }} \,
\label{vaff}
\end{eqnarray}
 where $\alpha=1$ or $2$. For symmetric and asymmetric ionic solutions, without colloids,
the PB theory also predicts  that, at large $r$, the effective interactions 
between the ions are Yukawa potentials similar to $v^{eff}_{0\, \alpha} (r)$.
This result, for instance, has been verified by computing  $v^{m}_{\alpha  \, \beta} (r)$ 
from the correlation functions $g_{\alpha  \, \beta} (r)$ determined by MC simulations 
for systems of charged hard spheres with equal diameters and  
densities of $\sim1$M \cite{Uland}. Similarly, an analytical expression of $v^{m}_{\alpha \, \beta} (r)$, 
with the same functional  dependence at large $r$ as $v^{eff}_{0\, \alpha}$, 
will be found from the correlation functions computed 
from the HNC OZ equations (\ref{dipolu}) and (\ref{dipolb}),
when the singularity nearest
to the real axis of the function $1/{\bar D}_0(k)$ 
is a simple pole located at $k = i\kappa_p$ \cite{Kjella,Ulande}.

Numerical methods have been derived for the determination of the zeros of ${\bar D}_0(k)$
for complex values of $k$ \cite{Evan,Leot,Dijks,Tutc,Grod}. They need the calculation of integrals having 
the typical expression 
 \begin{eqnarray} 
   I_p (k_0,k_1)= 4 \pi \int_0^\infty  \, r^2 \, c^{sr}_{\alpha,\beta}(r)
    {\rm sinh} (k_0 r) \cos (k_1 r) dr
 \label{cshr}
 \end{eqnarray}
where $c^{sr}_{\alpha,\beta}(r)= c_{\alpha,\beta}(r)+ l_B z_a z_b /r$  and $k_1$ and $k_0$
 are the real and imaginary parts of $k$. In practice, such integrals are not easily estimated  
 when the upper limit of the numerical integration on $r$ is $\sim$ 500 times 
 the radius of the ions.\par

 \begin{figure}
\includegraphics{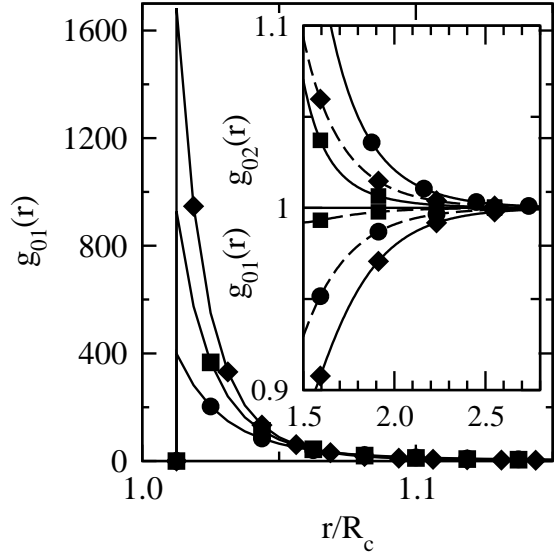}
\caption{\label{fig:epsarte} At $\rho_0=0$, $\rho_v=0.001$ M and ionic charges 
$z_1=-2,\,z_2=1$, variation of the functions $g_{01}(r)$ near
$r\simeq R_c$ : $z_0=1200$ solid line and filled circles,  
$z_0=1800$ solid line and filled squares, $z_0=2400$ solid line 
and filled diamonds.
Insert : for $1.5R_c<r<3R_c$, $g_{01}(r)$ (solid line and
filled circles $z_0=1200$, squares $z_0=1800$ and diamonds $z_0=2400$) 
and $g_{02}(r)$ (dashed line and filled circles $z_0=1200$,
 squares $z_0=1800$ and diamonds $z_0=2400$). }
\end{figure}

   This calculation requires at least an accuracy of the order of $10^{-8}-10^{-10}$ on 
  the $c^{sr}_{\alpha,\beta}(r)$ functions at large $r$ which is not  achieved
  with the present scheme used to solve the OZ equations. Therefore,
  we  analyse  the asymptotic behaviors  of the $g_{\alpha,\beta}(r)$ at
  infinite dilution by supposing that the  pole of $1/{\bar D}_0(k)$, 
  closest to the real axis,
  is effectively on the imaginary axis. With this hypothesis,
   $\beta v^{m}_{\alpha \, \beta} (r)$
  should be equal to $A_{\alpha \, \beta} \exp(-\kappa_p r)/r$ with $\kappa_p$
  identical for all the functions $v^{m}_{\alpha \, \beta} (r)$. 
  The determination of $A_{\alpha \, \beta} $
  and $\kappa_p$ is obtained from a fit of $\log\{g_{\alpha \, \beta} (r)\}$ 
  by $A_{\alpha \, \beta} \exp(-\kappa_p r)/r$, for $r$ sufficiently large. 
  However, the domains of large $r$ values where the functions 
  $\vert \log\{g_{\alpha \, \beta} (r)\}\vert$ are smaller than $10^{-6}$ 
  must be  excluded due to the numerical uncertainties on the 
  solutions of the OZ equations. The domains of $r$ values, 
  used in the fits, correspond to the distances for which $10^{-4}<$ 
  $\vert \, \log\{g_{\alpha \, \beta} (r)\} \, \vert$ $<10^{-1}$ where the numerical 
  uncertainties are two orders of magnitude smaller than the values of
  the fitted functions.\par

   Examples of curve fittings of  $\log\{g_{\alpha \, \beta} (r)\}$ 
  are given in Figs. 1 and  2 ; 
  these figures show unambiguously that these functions are accurately 
  described by the presupposed analytical form  $A_{\alpha \, \beta} \exp(-\kappa_p r)/r$.
   For a given ionic density, the values of $\kappa_p$ obtained from the curve fitting 
   of the functions $\log\{g_{0\,1} (r)\}$ and $\log\{g_{0\,2} (r)\}$ are independent of $z_0$
   when the values of $z_0$ vary from 1 to 4000 and identical to those determined by 
   the fit of  $\log\{g_{1\,1} (r)\}$, $\log\{g_{1\,2} (r)\}$ and $\log\{g_{2\,2} (r)\}$.
   At $\rho_v=0.1$, $0.01$ and $0.001$ and ionic charges ($z_1=-1,\,z_2=1$),
   ($z_1=-2,\,z_2=1$) and ($z_1=-1,\,z_2=2$), the values for $\kappa_p$ differ 
   by a few percent from $\kappa_D$ computed 
   for the same ionic concentrations and charges. At these weak packing
   fractions of the ions, the  differences between $\kappa_p$ and $\kappa_D$
   result from  contributions associated with the finite ionic density, 
   $\kappa_p$ being expected to be equal to $\kappa_D$ only in the limit of very 
   low densities where the functions $c^{sr}_{\alpha,\beta}(r)$ go to zero.\par

    In the HNC theory, the description of $g_{0 \,1} (r)$ and $g_{0 \,2} (r)$ at large $r$ 
    by a Yukawa function is in agreement with the expressions of the  PB 
    theory for the effective interactions  between ions and colloids
    at infinite dilution. It is thus possible to compare the
    values of the parameter $Z_{eff}$ recently computed from PB theory with those
    deduced from the values of the parameters $A_{0 \, 1}$ and $A_{0 \, 2}$
    by writing  $A_{0 \, \alpha}$ in a form identical to that derived from PB theory :
   \begin{eqnarray} 
 A_{0 \, \alpha} = Z^{HNC}_{eff} 
 z_\alpha l_B {\frac {\exp(\kappa_p \,  R_c) }
 { (1+\kappa_p R_c)}} \, .
 \label{zaaa}
 \end{eqnarray}

  Obviously, this expression does not result from a boundary condition similar to that satisfied
  by the solution of the PB equation on the colloid surface and can be considered  as a definition 
  of $Z_{eff}^{HNC}$. The possibility to write $A_{0 \, \alpha}$ in such a form is partly confirmed 
  by the fact that $A_{0 \, 1}/A_{0 \, 2}=z_1/z_2$ within the limits of the accuracy of the fit.\par 
  Recently PB expressions of $Z_{eff}$ have been derived. The most simple expression 
  \cite{triz1}, valid in the limit $z_0$ and $R_c /\sigma_i$ $\rightarrow \infty$,  is 
  \begin{eqnarray} 
  Z_{eff}^{a} = \frac{4 R_c}{l_B} ( 1 + \kappa_D \,  R_c) \, .
 \label{za}
 \end{eqnarray}

   An expression, a priori valid for all values of $z_0$ and $R_c$ 
   when, at infinite dilution, the colloids  are
   immersed in low density ionic solutions of monovalent ions,  
   has been derived \cite{triz2} by using
   the  asymptotically exact solution of the PB equation \cite{Shke}
 \begin{eqnarray} 
  Z_{eff}^{b} = \frac{4 R^2_c}{l_B} \, t(x) + 
  \frac{4 R_c}{l_B}\Big( 5- \frac{t^4(x)+3}{t^2(x)+1} \Big)\, t(x) \, .
 \label{zb}
 \end{eqnarray}
 where $x = z_0 l_B/(R_c \, (2 \kappa_D R_c +2))$ and $t(x) =(\sqrt{(1+x^2)} -1)/x$.
 This derivation  has been recently extended 
 to the cases of ion mixtures  with asymmetric charges 
 $z_1=-2$, $z_2=1$ and $z_1=-1$, $z_2=2$ \cite{triz3}. 
 It gives $Z_{eff}^{c}=z_0 \, f(x)$ where
 $f(x)$ is an intricate function of $x={z_0 \, l_B}/{\kappa_D \, R^2_c}$,
 depending on the charge asymmetry of the ions. The main results of these
 theoretical estimates are  that, for $z_0 = 1-100$,  $Z_{eff} \simeq z_0$ and, for
 $z_0 > 1000$,  $Z_{eff} $ goes to a constant value of the order of $Z_{eff}^{a}$.\par

 \begin{figure}
\includegraphics{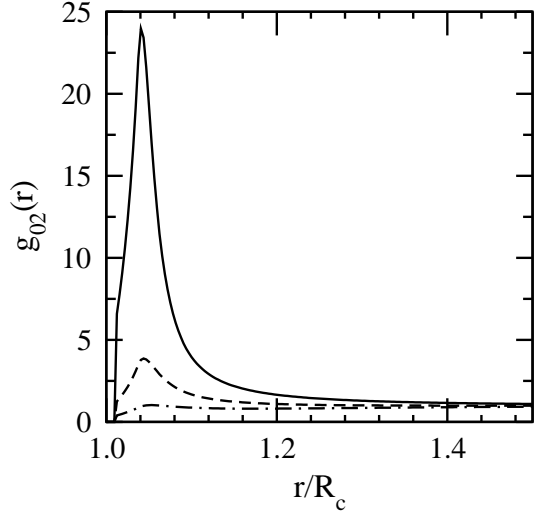}
\caption{\label{fig:epsartf} At $\rho_0=0$, $\rho_v=0.001$ M and ionic charges 
$z_1=-2,\,z_2=1$, variation of the functions $g_{02}(r)$ near
$r\simeq R_c$ : $z_0=1200$ dash-dotted line,  
$z_0=1800$ dashed line, $z_0=2400$ solid line.  }
\end{figure}

 From the fits of the functions $\log\{g_{\alpha\, \beta} (r)\}$, similar in accuracy to those 
 plotted in Figs. 1 and  2, we have computed the values of $\kappa_p$ and $Z_{eff}^{HNC}$ 
 resulting from the HNC OZ theory (cf. Eq. (\ref{zaaa})). At $\rho_v= 0.001$ M, 
 $\kappa_p R_c$ and $\kappa_D R_c$ are respectively equal to $2.089$  and $2.087$, 
 at $\rho_v= 0.01$ M equal to $6.62$ and $6.60$, and at $\rho_v= 0.1$
 they differ by 5\% and are equal to $21.8$ and $20.8$. Fig. 2 shows 
 that for $r > 2 R_c$, $g_{0\, 1} (r)$ and $g_{0\, 2} (r)$ are symmetric with respect to $1$,
 and Fig. 3 that the PB and HNC estimates $Z^{b}_{eff}$ and $Z^{HNC}_{eff}$ 
 are in good agreement at the ionic 
 concentrations of $0.001$ and $0.01$ M when $z_0 < 2000$, in particular the two estimates 
 reach a maximal value near $z_0 \simeq 2000$. For larger values of $z_0$ the PB and HNC 
 estimates $Z_{eff}^{b}$ and $Z_{eff}^{HNC}$ differ systematically : $Z_{eff}^{b}$ stays 
 constant whereas $Z_{eff}^{HNC}$ decreases.\par

   At $\rho_v= 0.1$ M, in the considered range 
 of $z_0$, the PB values of $Z_{eff}$ do not reach their asymptotic value.
 The HNC estimates of $Z_{eff}$ would have values exceeding largely $z_0$ if $Z_{eff}^{HNC}$ 
 was calculated from Eq. (\ref{zaaa}) by using $\kappa_D$ and not $\kappa_p$. 
 This result seems to be an indication that the finite concentration effects 
 of co-ions and counter-ions cannot be neglected 
 when the ionic concentrations are of the order of or larger than $0.1$ M.\par

 For the charge asymmetric ionic mixtures of $\rho_v= 0.001$ M, $z_1=-2, \, z_2=1$
 and $z_1=-1, \, z_2=2$, Fig. 4 presents a comparison between  
 $Z^{c}_{eff}$ and $Z_{eff}^{HNC}$. For the case $z_1=-1, \, z_2=2$, 
 the qualitative agreement is excellent
 and the quantitative difference is not larger than 10\%. 
 The two estimates of $Z_{eff}$
 are almost constant when $z_0$  $> 1000$. Fig. 5 shows that, in agreement
 with the fact that $Z_{eff}^{HNC}$ is quasi constant for these 
 values of $z_0$, $g_{0\, 1} (r)$ and $g_{0\, 2} (r)$ vary only near 
 $r\, \simeq \, R_c$ and stay almost unchanged when $r\, > 2\, R_c$.
 In the case $z_1=-2, \, z_2=1$, $Z^{c}_{eff}$ and $Z^{HNC}_{eff}$ differ for
 $z_0 > 500$, $Z_{eff}^{c}$ is constant but $Z^{HNC}_{eff}$ decreases and becomes 
 negative near $z_0=1800$. Fig. 6 shows the variation of $g_{0\, 1} (r)$ and 
 $g_{0\, 2} (r)$ which, taking into account the ratio $z_1/z_2$ and the sign
 inversion of $Z_{eff}^{HNC}$,
 retain the correct symmetry with respect to 1 when $r > 2 \, R_c$.\par
 \begin{figure}
\includegraphics{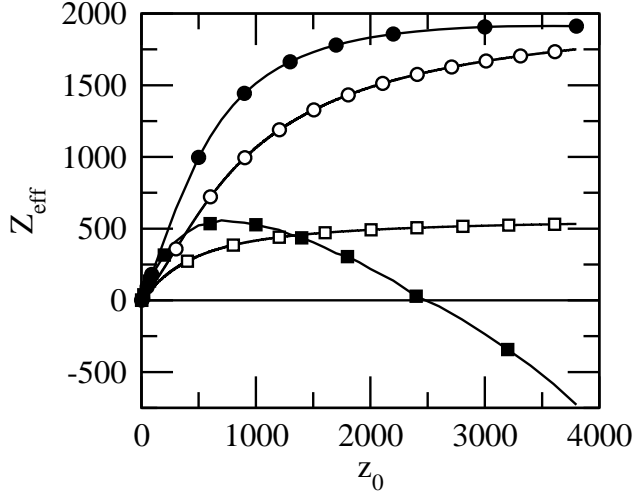}
\caption{\label{fig:epsartg}At $\rho_0=0$, PB and HNC estimates 
of $Z_{eff}$ versus $z_0$ 
for ionic concentrations $\rho_v=0.01$ M, ionic charges $z_1=-1,\,z_2=2$,
 : solid line and white circles PB, solid line and filled circles HNC, 
and  $z_1=-2,\,z_2=1$  
 : solid line and white squares PB, solid line and filled squares HNC.}
\end{figure} 
  For $r <  2 \, R_c$, the main feature is the appearance of
 a peak in  $g_{0\, 2} (r)$ (cf. Fig. 7) near $r \simeq R_c+1.5\sigma_i$ characterizing  
 an overscreening of the colloid charge for large $z_0$ by the negative ions 
 which allows the positive ions to be located close to the colloid surface.
 This overscreening seems to be at the origin the sign inversion 
 of the colloid-ion potential
 of mean force, which is not obtained in the PB theory.\par 
 A similar behavior is found
 for $\rho_v= 0.01$ as shown by the comparison between $Z^{c}_{eff}$ and $Z^{HNC}_{eff}$
 presented in Fig. 8. At $\rho_v= 0.1$, $z_1=-2$, $z_2=1$ ($\kappa_p R_c > 25 $), 
 Fig. 9 shows 
 that $g_{0\, 2} (r)$ exhibits a peak at $r \simeq  R_c+1.5\sigma_i$ for  $z_0=3800$
 which indicates an overscreening of the colloid charge. However the 
  large value  of $\kappa_p R_c$ 
 implies a very steep decrease of the correlation functions and 
 the range of $r$ values where a significant fit of $g_{0\, 1} (r)$
 and $g_{0\, 2} (r)$ is possible, is very small ($\simeq 0.1 R_c$) and 
 does not allow for an estimate of $Z^{HNC}_{eff}$.\par

\begin{figure}
\includegraphics{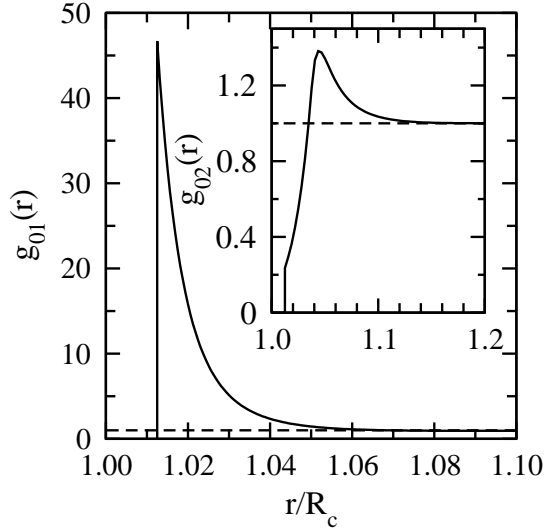}
\caption{\label{fig:epsarth} At $\rho_0=0$, $\rho_v=0.1$ M  $z_1=-2,\,z_2=1$  
and $z_0=3800$ : ion-colloid correlation functions 
$g_{01}(r)$ and $g_{02}(r)$.}
\end{figure}

\begin{table}
\caption{\label{tab:table1}
Colloid packing fraction : $\eta_c$, density of the ion species 
with the highest valence :
$\rho_v $, ion charges : $z_1 $ and $z_2$, 
maximum colloid charge : $z_0^{\rm max}$,
$\kappa_D$ and  $\kappa^c_D$ : inverse screening Debye 
length without and with the 
colloid charge contribution. 
The last line corresponds 
to a binary ion-colloid mixture.}

\begin{tabular}{|c|c|c|c|c|c|c|}
\hline
\ \ \ \ $\eta_c$ \ \ \ \ & \ \ \ $\rho_v \ M$ \ \ \ &
\ \ $z_1 $ \ \ & \ \ $z_2$\ \  &\  $z_0^{\rm max}$ &\  $\kappa_D\, R_c$ \ 
&\  $\kappa^c_D\, R_c$ \  \\ \hline
 0.0001 &  0.001 & -1  & 1 &  450 &  2.09  & 2.10  \\  \hline
 0.0001 &  0.001 & -2  & 1 &  270 &  2.55  & 2.55  \\  \hline
 0.0001 &  0.001 & -1  & 2 &  630 &  3.61  & 3.61  \\  \hline
 0.001 &  0.001 & -1  & 1 &  300  &  2.09  & 2.10  \\  \hline
 0.001 &  0.001 & -2  & 1 &  260  &  2.57  & 2.57  \\  \hline
 0.001 &  0.001 & -1  & 2 &  465  &  3.62  & 3.62  \\  \hline
 0.01 &  0.001 & -2  & 1 &  189   &  2.67  & 2.67  \\  \hline 
  0.1 &  0.003 & -1  & 1 &  109   &  2.35  & 2.59  \\  \hline
\hline  
  0.1 &        & -1  &   &   130   &     & 1.67  \\  \hline
 \end{tabular}

\end{table}

\section{Finite concentration}

 The  {\em binary} mixture composed of colloids and counterions with $z_1=-1$, 
 without 
 added salt, has been well studied in the literature \cite{Bel,Bell}
 and we briefly present and discuss
 the HNC results obtained for the case of a colloid-counterion size ratio equal to 80 
 at a colloid packing fraction $\eta_c=\pi \rho_c \sigma_c^3/6=0.1$. For this mixture,
 a numerical solution of the HNC OZ equations is only  found when  
 $z_0<130$. This limitation on the  $z_0$ values is similar to that obtained 
 in Refs. \cite{Bell,Lai,Anta}. At $z_0=130$, using the previously adopted
 values of ion and colloid radii, the ionic concentration is $\simeq 0.0064$ M
 and the ionic packing fraction $\simeq 0.0000253$. Even at low values of $z_0$
 the asymptotic behavior of the three correlation functions corresponds
 to damped oscillations (cf. Fig. 10) indicating 
 that, in the OZ equations (cf. Eq. (\ref{dipolu})),
 the asymptotic behavior is determined by
 poles of $1/{\bar D}_0(k)$ symmetric with 
 respect to the imaginary axis. A fit of the pair correlation functions
 at $z_0=130$ for $8 R_c<r<25R_c$ (cf. above and below for a detailed discussion 
 of the fitting procedure) shows that these functions oscillate 
 with the same period equal to $1.95R_c$ (cf. Fig 11) and have the  same damping
 constant $\kappa_p \, R_c=0.24$ about 6 times smaller
 than the inverse screening Debye length $\kappa_D \, R_c=1.67$ 
 (cf. Table 1).\par
 
\begin{figure}
\includegraphics{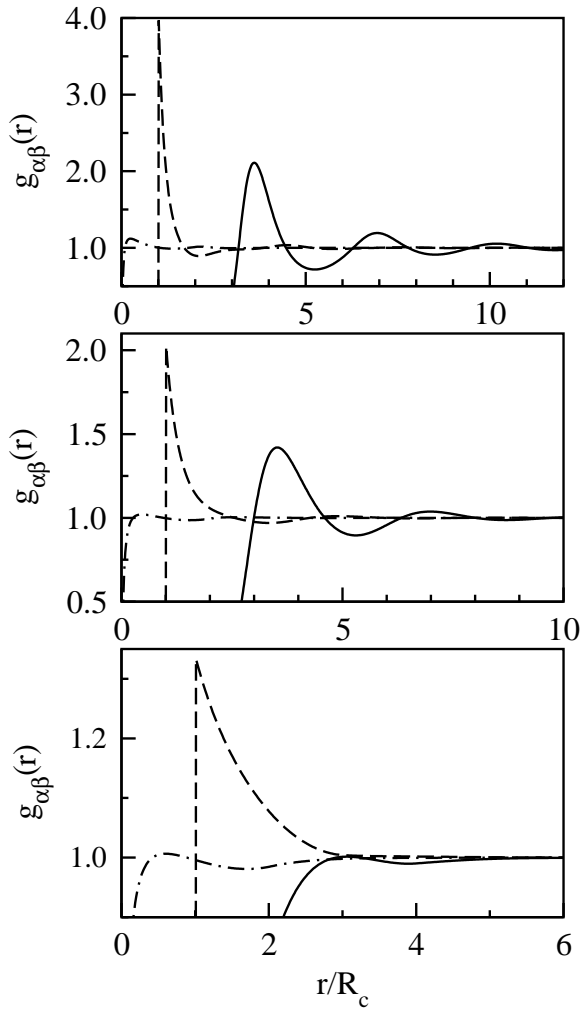}
\caption{\label{fig:epsarti} At $\eta_c=0.1$, $\rho_2=0$ and   $z_1=-1$  
colloid-colloid $g_{00}(r)$ (solid line), ion-colloid 
$g_{01}(r)$ (dashed line) and ion-ion $g_{11}(r)$ (dash-dotted line) 
correlation functions, from bottom to top $z_0=10$, $z_0=50$ and $z_0=130$. }
\end{figure} 

 In Fig. 10, it is seen from
 $g_{0\, 1} (r)$ that the screening of the colloid charges occurs 
 over a distance which varies between 1.5 and 0.6 $R_c$
 when $z_0$ increases from 1 to 130. For these values of $z_0$,
 the range of the screening effects stays sufficiently large 
 so that the coulombic repulsion shifts the main peak of 
 $g_{0\, 0} (r)$ from 2$R_c$ to 3.8$R_c$. This displacement
 induces a reinforcement of the excluded volume effects between the colloids. 
 For $z_0\simeq 50-130$, $g_{0\, 0} (r)$ has the typical form  
 of the pair correlation function of a dense monoatomic fluid.  
 It seems, although difficult to establish numerically,
 that this increase of the excluded volume effects is the main
 cause of the instability of the HNC OZ equations at $z_0> 130$.
 
\begin{figure}
\includegraphics{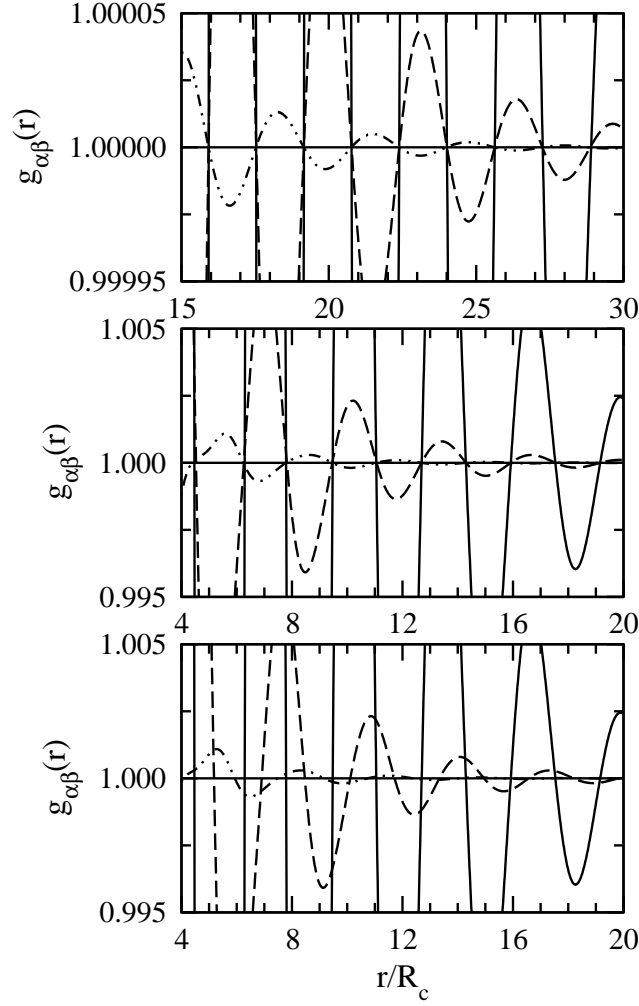}
\caption{\label{fig:epsartj} 
At $\eta_c=0.1$, $\rho_2=0$, $z_1=-1$ and $z_0=130$, 
: colloid-colloid $g_{00}(r)$ (solid line), ion-colloid 
$g_{01}(r)$ (dashed line) and ion-ion $g_{11}(r)$ (dash-dotted line) 
correlation functions. From bottom to top : $g_{00}(r)$, $g_{01}(r)$ 
and $g_{11}(r)$, and, after an appropriated shift of 
$g_{01}(r)$ and $g_{11}(r)$ on $r$ equal to $d^p_{01}$ and  $d^p_{02}$
(cf. Eq. [\ref{dedc}]), for $4< r/R_c < 20$
and  $15< r/R_c < 30$  close views
showing the identity of the oscillation period 
of the three correlation functions (the maxima and minima
of the oscillations of $g_{00}(r)$ and $g_{01}(r)$ can be truncated).  }
\end{figure}

 The {\em ternary} solutions of colloids, positive and negative ions have
 been considered for different colloid packing fractions, ionic
 concentrations and colloid charges summarized in Table 1. In these mixtures 
 for given densities of the colloids and the positive ions, the density 
 of the negative ions is fixed by the electroneutrality condition. 
 In Table 1, $z_0^{max}$ denotes the maximum value of $z_0$ at
 which the HNC equations can be solved by using the previously described numerical 
 iterative scheme. 
 In the mixtures, the asymptotic behaviors of the correlation functions 
 $g_{\alpha \, \beta} (r)$  can be analysed similarly to the 
 infinite dilution case, since a typical OZ equation for a correlation function
 has the form
 \begin{equation} 
      {\bar h}_{01}(k) = \frac{({\bar c}_{01}(k)+x_2\, {\bar c}_{02}(k)\, {\bar c}_{12}(k)
	-x_2\, {\bar c}_{22}(k)\, {\bar c}_{01}(k))}{{\bar D}(k)}
 \label{dipt}
 \end{equation} 
where the expression of ${\bar D}(k)$, identical for all ${\bar h}_{\alpha \, \beta}(k) $, 
is 
\begin{eqnarray}
 {\bar  D}(k) & = & (1-x_2\, {\bar c}_{22}(k)-x_1\, {\bar c}_{11}(k)-x_0\, {\bar c}_{00}(k)\nonumber \\
 &+&x_1\, x_2\, {\bar c}_{11}(k)\, {\bar c}_{22}(k)  
      -   x_1\,  x_2\, {\bar c}_{12}(k)^2 \nonumber \\
    & - & \, x_0\, x_1 \, {\bar c}_{01}(k)^2-x_0\,  x_2 \, {\bar c}_{02}(k)^2  \nonumber \\
  &+ & x_0\, x_1\, {\bar c}_{11}(k)\,  {\bar c}_{00}(k)  
      -   x_0\, x_1\,  x_2\, {\bar c}_{11}(k)\, {\bar c}_{00}(k)\, {\bar c}_{22}(k) \nonumber \\
    & + & x_0\,  x_1\, x_2 {\bar c}_{11}(k)\, {\bar c}_{02}(k)^2  \nonumber \\ 
   & - &  2\, x_0\,  x_1\, x_2\,{\bar c}_{01}(k)\, {\bar c}_{12}(k)\, {\bar c}_{02}(k) \nonumber \\
    & + &  x_0\,  x_2\,{\bar c}_{00}(k)\, {\bar c}_{22}(k) 
     +  x_0\, x_1\, x_2\, {\bar c}_{01}(k)^2\, {\bar c}_{22}(k) \nonumber \\
     & + &  \, x_0\, x_1\, x_2\, {\bar c}_{12}(k)^2\, {\bar c}_{00}(k))  \, 
 \label{dipf}
 \end{eqnarray} 
 where $x_0=\rho_0/\rho$.\par
\begin{figure}
\includegraphics{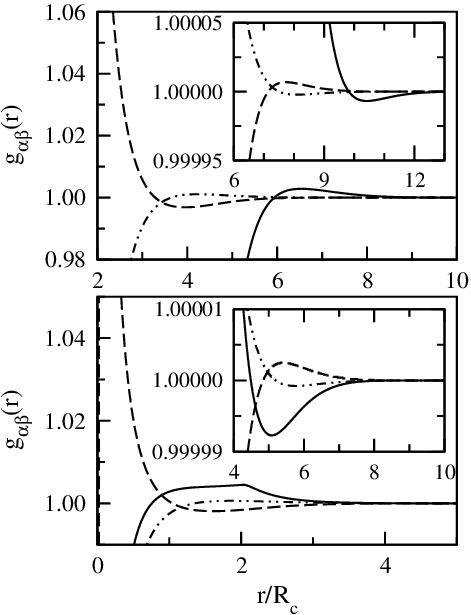}
\caption{\label{fig:epsartk} At $\eta_c=0.001$, 
$z_1=-1,\; z_2=1$, $\rho_v=0.001$ M
and $z_0=300$, {\it bottom} : correlation functions $g_{11}(r)$ (solid line), 
$g_{12}(r)$ (dashed line) and  $g_{22}(r)$ (dash-dotted line), 
insert : close view  for $r > 4 R_c$, and 
{\it top} : $g_{00}(r)$ (solid line), 
$g_{01}(r)$ (dashed line) and  $g_{02}(r)$ (dash-dotted line),
insert : close view  for $r > 6 R_c$.}
\end{figure} 
  
Clearly, the zeros of ${\bar D}(k)$, if they exist at complex values of $k$ near 
the real axis, determine the decrease of all the functions $g_{\alpha \, \beta} (r)$.
Since these functions are real, the most simple locations of these zeros 
correspond to values of $k$  on the imaginary axis, $k=i\kappa_p$, 
giving at large $r$ an exponential decrease of 
$g_{\alpha \, \beta} (r)\simeq 1+A_{\alpha \, \beta} \exp(-\kappa_p r)/r$, 
or to values of $k$ symmetric with respect to the imaginary axis,
 $k=k_0+i\kappa_p$ and $k=-k_0+i\kappa_p$, giving an oscillatory 
exponential damped decrease of $g_{\alpha \, \beta} (r) \simeq 
1+A_{\alpha \, \beta} \exp(-\kappa_p r) \cos(k_0 r+ d_{\alpha \, \beta})/r$. 
In the first case, at large $r$ all correlation functions are exponentially decreasing,
in the second case all the functions $g_{\alpha \, \beta} (r)$ oscillate 
with an identical period and damping. Contrary, to the infinite dilution case
where, for the considered ionic solutions, an exponential decrease  was always
obtained,
both exponential and oscillatory damped decreases are found
for finite colloid concentrations.\par 

\begin{figure}
\includegraphics{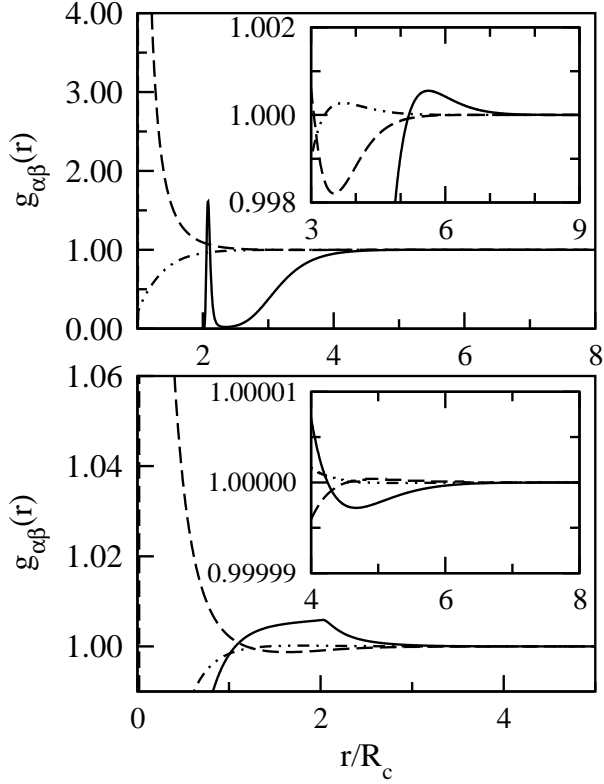}
\caption{\label{fig:epsartl}At $\eta_c=0.001$, $z_1=-2,\; z_2=1$, $\rho_v=0.001$ M
and $z_0=259$, {\it bottom} : correlation functions $g_{11}(r)$ (solid line), 
$g_{12}(r)$ (dashed line) and  $g_{22}(r)$ (dash-dotted line),
insert : close view  for $r > 4 R_c$, and {\it top} : $g_{00}(r)$ (solid line), 
$g_{01}(r)$ (dashed line) and  $g_{02}(r)$ (dash-dotted line),
insert : close view  for $r > 3 R_c$ (the scale of $y$-axis precludes
to identify the maxima and minima for $r> 6R_c$). }
\end{figure} 

For instance, in the case
of a symmetric ionic solution ($z_1=-1,\; z_2=1$), at $\rho_v=0.001$M and 
 $\eta_c=0.001$, the OZ equations can be solved in
the domain $1<z_0<300$. For  $z_0<100$,
 $g_{0\, 0} (r)$ increases monotonically
from 0, at short distance, towards $1$ at large $r$. 
Above this  value of $z_0$ (cf. Fig. 12), $g_{0\, 0} (r)$
presents, at $r \simeq 6-7 R_c$, a weak broad peak, larger than 1, then,
at $r \simeq 10 R_c$, a  minimum smaller than $1$, and finally, at large $r$,  
 increases monotically towards $1$.
A similar behavior is obtained for $g_{1\, 1} (r)$ ; the peak and minimum,
which exist also for $z_0>100$, are located at $r \simeq 1-2 R_c$
and at $r \simeq 5-6 R_c$, respectively. 
At larger $r$, $g_{1\, 1} (r)$
increases monotically towards $1$.  The function $g_{2\, 2} (r)$, different from
$g_{1\, 1} (r)$ due to the finite density of the colloids, 
 presents   a peak and minimum for
$z_0>100$ and a final monotonic increase towards 1 at large $r$. The function 
$g_{1\, 2} (r)$, has a peak near $r=\sigma_i$, obviously due to the attraction
between ions of opposite charges. For $r> 1.5 R_c$ and 
$z_0>100$, the monotonic decrease of this peak towards $1$ is modified,
$g_{1\, 2} (r)$ has a minimum, smaller than 1, at $r \simeq 2-3R_c$,
followed by a local maximum at $r=5R_c$, and  for larger $r$ decreases
monotonically towards $1$.\par

 The functions 
$g_{0\, 1} (r)$ and $g_{0\, 2} (r)$ when $z_0<100$
have the expected behavior at large $r$ characterized, respectively,
 by a monotonic decrease for
$g_{0\, 1} (r)$ and increase for $g_{0\, 2} (r)$ towards $1$.
But, above $z_0>100$, $g_{0\, 1} (r)$ ($g_{0\, 2} (r)$)
has, a local minimum (maximum) at $r\simeq4R_c$,
followed at $r\simeq7R_c$ by a a local maximum (minimum)
and at large $r$ decreases (increases) towards $1$.\par
In summary, for low values of $z_0$, the asymptotic behavior of
the correlation functions seems exponential, 
as in the case of infinite dilution of the colloids.
For a large value of $z_0$, 
it presents oscillations, although, within the limit of numerical accuracy,
only one or two oscillations can be observed unambiguously. \par
\begin{figure}
\includegraphics{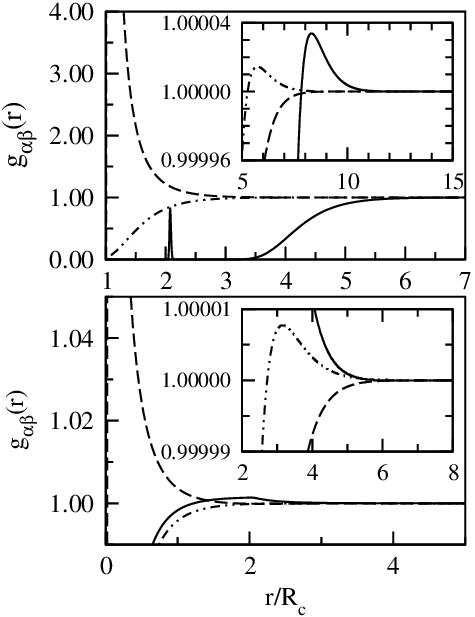}
\caption{\label{fig:epsartn}  At $\eta_c=0.0001$, 
$z_1=-1,\; z_2=1$, $\rho_v=0.001$ M
and $z_0=460$, {\it bottom} : correlation functions $g_{11}(r)$ (solid line), 
$g_{12}(r)$ (dashed line) and  $g_{22}(r)$ (dash-dotted line),
insert : close view  for $r > 2 R_c$,
 and {\it top} : $g_{00}(r)$ (solid line), 
$g_{01}(r)$ (dashed line) and  $g_{02}(r)$ (dash-dotted line),
insert : close view  for $r > 5 R_c$. }
\end{figure}

When $z_1=-2$ and $z_2=1$  or $z_1=-1$ and $z_2=2$ ,
similar analysis of the asymptotic behavior of the correlation functions
can be made with identical conclusions, since in these cases, 
the HNC equations can be solved for $z_0<260$ and $z_0<465$, respectively
(cf. Table 1). The main remark
concerns the system with $z_1=-2$ and $z_2=1$, where the onset of a peak 
in $g_{0\, 0} (r)$ at $r \simeq 2.08 R_c$ for $z_0>250$, 
indicates that the screening of the colloid charges by negative ions
has a range sufficiently short for two colloids
to be separated by a distance of one or a few ion diameters (cf. Fig.  13).\par

At the packing fraction $\eta_c=0.0001$ and $\rho_v=0.001$ M, 
the HNC equations can be solved numerically  for symmetric
($z_1=-1$ and $z_2=1$) and asymmetric ($z_1=-1$ and $z_2=2$,
$z_1=-2$ and $z_2=1$) ions up to $z_0 < 460$, $z_0 < 630$ 
and $z_0 < 270$, respectively. 
The main difference, compared to the case of
the colloid packing fraction of $\eta_c=0.001$, 
is that, for $z_0>100$, a peak or  minimum appears 
at $r > 2R_c$ in the ion-ion correlation functions
and at $r > 4R_c$ or $r > 6R_c$ in the ion-colloid or colloid-colloid 
correlation functions. Beyond this peak or minimum, the correlation
functions go monotonically towards $1$
within the numerical accuracy of the HNC solutions.
For the three considered cases of ionic charges, near 
$z_0^{max}$, a peak appears in $g_{0\, 0} (r)$ 
for $r \simeq 2.08 R_c$, indicating that the screening 
of the colloid charges allows a close approach between 
colloids (cf. Fig.  14).\par

In the other studied mixtures at $\eta_c=0.01$,
and $\eta_c=0.1$ with symmetric and asymmetric ionic 
charges, the correlation functions have asymptotic behaviors  
qualitatively similar to those found for lower
colloid packing fractions. The main quantitative difference
is that the non-monotonic decay of the correlations appears
at low colloid charges $z_0\simeq 10$. For the
colloid packing fraction $\eta_c=0.1$, 
the correlation functions have an oscillatory
behavior at large $r$ for $z_0>20$.
At this value of $\eta_c$ and $z_0>70-80$,
several oscillations can be identified at large $r$
with amplitudes larger than the numerical error.\par 
 
 From the results obtained at a finite density of the colloids,
it appears that  the asymptotic behavior  of
the correlations functions, as computed from the
HNC OZ equations, is compatible with the hypothesis 
that it is determined by the poles of $1/{\bar D}(k)$ 
as only monotonic or oscillatory behaviors have been
found. It is worth  remarking  that, since 
the type of asymptotic decrease is the same for all the correlation functions,
it should be sufficient to consider only one of
the correlation functions for determining this behavior, for instance 
$g_{0 \, 0} (r)$. However, as in the case of the infinite 
dilution colloid system, such a determination, 
due to the present numerical accuracy  on $g_{0\, 0} (r)$
can be performed  only in the range of $r$ for which  
$\vert g_{0\, 0} (r)-1\vert>10^{-6}$. A similar remark
obviously applies to all the other correlations functions.
With the hypothesis that the asymptotic behavior of
the functions $g_{\alpha \, \beta} (r)$ is determined by 
simple poles near the real axis, and considering
only the two nearest poles to the real axis,
a function $g_{\alpha \, \beta} (r)$ writes \cite{Uland,Kjella,Ulande}
\begin{eqnarray} 
g_{\alpha \, \beta} (r)&\simeq& 1+A_{\alpha \, \beta}^p \exp(-\kappa_p r)
\cos(k^0_p r+ d_{\alpha \, \beta}^p)/r \nonumber \\
 &+& A_{\alpha \, \beta}^o \exp(-\kappa_o r) \cos(k^0_o r+ d_{\alpha \, \beta}^o)/r\, .
 \label{dedc}
 \end{eqnarray}
 The relative values of  $\kappa_p$ and $\kappa_o$
determine the dominant contribution to the asymptotic behavior.
When, for instance, $\kappa_p<<\kappa_o$, and
 $k^0_p$ and $d_{\alpha \, \beta}^p$  equal to  or different from zero,  
 the decrease at large $r$ is exponential or damped oscillatory ;
 when $\kappa_p \simeq \kappa_o$, obviously the type of decrease 
 can be difficult to identify
from the numerical values of $g_{\alpha \, \beta} (r)$, in particular
when  $A_{\alpha \, \beta}^p<<A_{\alpha \, \beta}^o$. Figure 15
shows the behavior of  the functions $g_{\alpha \, \beta} (r)$ for
the suspensions with $\eta_c=0.1$, $z_0=108$, $z_1=-1$, $z_2=1$ and $\rho_v=0.003$M.
All functions present damped oscillations for $r>6 R_c$, in particular, 
$g_{0\, 0} (r)$ which is above the level of numerical uncertainties until $r\simeq15R_c$.
This result indicates that, assuming $\kappa_p<\kappa_o$ and a value of
$A_{0 \, 0}^p$ estimated  from  Eq. (\ref{zaaa}),
the asymptotic behavior of $g_{0\, 0} (r)$ is not determined by 
the Debye screening since for $r\simeq 12R_c$, the factor 
$\exp(-\kappa_D r)$ is equal to $\simeq 10^{-19}$.
In Fig. 15, as expected, the oscillations in all correlation functions have
an identical period, although the numerical accuracy and 
short range contributions for $r\simeq 4-5 R_c$ can preclude 
an exact coincidence of the oscillations which, however, is realized
with an accuracy of $10^{-3}R_c$ for $r>5R_c$.\par 

\begin{figure}
\includegraphics{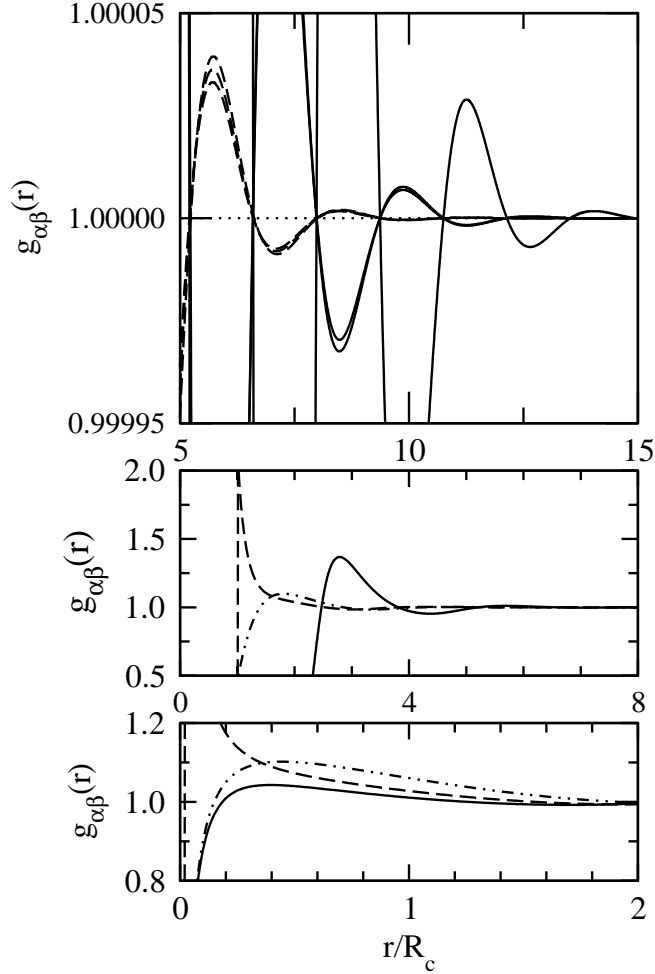}
\caption{\label{fig:epsartm}  At $\eta_c=0.1$, $z_1=-1,\; z_2=1$, $\rho_v=0.003$ M
and $z_0=108$, {\it bottom} : correlation functions $g_{11}(r)$ (solid line), 
$g_{12}(r)$ (dashed line) and  $g_{22}(r)$ (dash-dotted line), 
{\it middle} : $g_{00}(r)$ (solid line),
$g_{01}(r)$ (dashed line) and  $g_{02}(r)$ (dash-dotted line),
{\it top} : for $r > 5 R_c$, close view  showing the identity 
of the oscillation period of the six correlation functions, 
after an appropriate shift on $r$ (cf. Fig. 11) 
of the  ion-ion (dashed line) and
colloid-ion (solid line) correlation functions, 
(notice that the maxima and minima
of the oscillations of $g_{00}(r)$, $g_{01}(r)$ and $g_{02}(r)$ 
can be outside the limits of the figure).}
\end{figure} 

Another example where the asymptotic behavior does not seem to be
determined by the Debye screening, is found for $\eta_c=0.001$,
$z_0=300$, $z_1=-1$, $z_2=1$ and $\rho_v=0.001$ M. This conclusion
is obtained from the 
fact that $g_{0\, 0} (r)$ is significantly larger
than $10^{-6}$ for $r\simeq 10-12R_c$, a domain of $r$ where 
$\exp(-\kappa_D  r)\simeq 10^{-10}$ (cf. Table 1).
For this suspension, the asymptotic behavior of the correlations
seems well described by a superposition of two exponentials
($k^0_p=0$, $k^0_o=0$) where $ A_{\alpha \, \beta}^p$ and 
$A_{\alpha \, \beta}^o$ have  opposite signs (cf. Fig. 12). 

At $\eta_c=0.0001$, $z_0=460$, $z_1=-1$, $z_2=1$ and $\rho_v=0.001$M,
the  asymptotic behavior is described by one exponential with
a screening length $\kappa_p$ compatible with the value of the
Debye length, because the maximum of $g_{0\, 0} (r)$
exists at $r\simeq 8 R_c$ where $\exp(-\kappa_D  r)\simeq 10^{-7}$.
However it is worth remarking that the
signs of the $ A_{\alpha \, \beta}^p$ coefficients for the colloid-colloid
and ion-colloid functions are opposite to their expected
values ; for instance $ A_{0\, 0}^p$ is positive (cf. Fig. 14). Clearly, 
this fact seems to preclude the interpretation  of $A_{0\, 0}^p$
in terms of an effective charge which implies $ A_{0\, 0}^p \propto -\, Z^2_{eff}$ 
 (cf. Eq. (\ref{veff})).

For the considered suspensions, the OZ HNC 
equations predict  that the  asymptotic correlation of the 
two-body functions has several possible behaviors.
For $z_0>100$, the decrease of these correlations
at large $r$ generally does not correspond to that induced
by the Debye screening. Due to the large difference of 
size between colloids and ions, the colloid-colloid 
correlations, which have a range of several
colloid diameters, determine at large $r$ 
the behavior of all the correlations between the ions 
and the colloids and ions, as a consequence of
the exact OZ relations. In particular, when $z_0$ increases, 
the excluded volume effects between the colloids
are reinforced and  $g_{0\, 0} (r)$ presents oscillations
and looks like the pair correlation of a fluid of soft core
particles at low or moderate densities.

\section{Conclusion}

 At the infinite dilution of the colloids, 
 the ion-colloid pair correlation functions computed from
 the OZ HNC equations allows to define effective colloid-ion potentials 
 and colloid-ion charges ($Z_{eff}$). The results obtained for ionic
 solutions where the ion density varies between $0.001$M to $0.1$M
 are in fair agreement with the 
 effective potentials and charges obtained from PB theory when $z_0$ is
 smaller than 1000. For larger $z_0$, the PB
 and HNC values of $Z_{eff}$ differ. While
 the PB values  go  towards a 
 constant value both for symmetric and asymmetric
 ionic solutions, the HNC values
 decrease and, in the case of the
 asymmetric solution $z_1=-2$ and $z_2=1$,
 $Z_{eff}^{HNC}$ becomes negative. This last result seems
 to be due to an overscreening of the colloid
 charge. The HNC theory shows that, when the ionic concentration 
 is of the order of $0.1$ M, the corrections to the Debye screening length
 due to finite density effects are not negligible in the computation
 of $Z_{eff}$. 
 Although the difference between $\kappa_p$ and $\kappa_D$ is only
 5\%, it induces a factor of 3 on the value of
 $Z_{eff}$ when in Eq. (\ref{zaaa}) $\kappa_D$ is substituted
 to $\kappa_p$.
 
 At finite concentrations of the colloids, all the correlation functions
 present an identical asymptotic qualitative behavior. The 
 colloid-colloid correlations determine the decrease of this behavior.
 The analysis of the asymptotic behavior does not seem
 to allow, as in the case of infinite dilution, to define in
 a simple way a colloid-colloid effective potential or colloid 
 effective charge from $v^{m}_{\alpha \, \beta} (r)$. 
 The main reason is that, at large $r$, the colloid-colloid
 correlations, for large $z_0$ values may be dominated
 by excluded volume effects similar to those existing
 in fluids of particles interacting by a hard or soft core 
 repulsive interaction. This result is in agreement with
 the fact that the range of the correlations is larger
 than that expected from the value of the Debye length 
 controlling the charge screening. Obviously, it leaves
 open the possibility, largely discussed in the literature,
 to define an effective colloid-colloid potential,
 able to reproduce adequately both screening and excluded
 effects from a relation different from $\log \{g_{0 \, 0} (r)\}$
 $=$ $-\beta v^{m}_{0 \, 0} (r)$ \cite{Biben,Clem,Gottw}.
 
 The HNC approximation has been proved reliable to describe the
 coulombic fluids by comparison with MC simulations.
 For the present model, such a test of the validity of HNC 
 approximation seems prevented by the difficulty of realizing
 an efficient MC sampling of the configurations 
 of colloid suspensions where the ratio of
 colloid and ion sizes is 80 and the colloid charges are 
 much larger than 100. However, if the HNC approximation 
 generally gives good quantitative results when it can be solved, 
 the stability limit of the numerical solutions can not be faithfully
 associated with a thermodynamic transition. Thus,
 at the colloid packing fraction $\eta_c=0.001$, the loss
 of stability of the numerical HNC solutions corresponding to
 the onset of a peak in $g_{0\, 0} (r)$ at $r\simeq 2.08$, 
 cannot unambiguously be interpreted as the formation of 
 colloid aggregates. 
 
From the present results and those published in the literature
\cite{Kusal,Kusali}
it seems possible to extend the use of the HNC OZ equations 
for studying colloidal suspensions with colloid-ion size ratio larger
than 80 and, also, to consider ionic  solutions 
with  a discrete solvent, at least 
for an infinite dilution of the colloids and to determine the influence 
of the solvent on the ion and solvent-colloid correlation functions 
and the values of $Z_{eff}$.\par  
\begin{acknowledgments}
The authors are indebted to Jean-Michel Caillol and Emmanuel Trizac  
for helpful discussions.
\end{acknowledgments}

\end{document}